\documentclass[12pt]{article}
\usepackage{axodraw}
\newcommand{\wbar}{{\bar {\phantom m}}\!\!\!\;\!\!{\bar {\phantom m}}
\!\!\!\!\!\!}
\begin{document}
\begin{center} {\Large \bf  Renormalization of the Fayet-Iliopoulos
Term\\[4mm] in Softly Broken SUSY Gauge Theories} \vspace{1cm}

{\large \bf D.I.Kazakov$^{\dag , \ \ddag}$ and
V.N.Velizhanin$^\dag$ }\vspace{0.7cm}

{\it $^\dag$ Bogoliubov Laboratory of Theoretical Physics, Joint
Institute for Nuclear Research, Dubna, Russia \\[0.2cm] and\\[0.2cm]
$^\ddag$ Institute for Theoretical and Experimental Physics,
Moscow, Russia}
\end{center}

\begin{abstract} It is shown that renormalization of the Fayet-Iliopoulos
term in a softly broken SUSY gauge theory, in full analogy with all
the other soft terms renormalizations, is completely defined in a
rigid or an unbroken theory. However, contrary to the other soft
renormalizations, there is no simple differential operator that
acts on the renormalization functions of a rigid theory and allows
one to get the renormalization of the F-I term. One needs an
analysis of the superfield diagrams and some additional diagram
calculations in components.
The method is illustrated by the four loop calculation of some part of
renormalization proportional to the soft scalar masses and the soft triple
couplings.
\end{abstract}\vspace{0.3cm}

\section{Introduction}
In our previous publications~\cite{AKK,K98,KV}, we gave a complete
set of the rules needed for writing down the RG equations for the
soft SUSY breaking terms in an arbitrary non-Abelian N=1 SUSY
gauge theory. Our main statement is that all the renormalizations
in a softly broken SUSY theory are completely defined by the
rigid, or unbroken, theory and may be evaluated by the use of
simple differential operators~\cite{Y,JJ,AKK,KV} or by expansion over
the Grassmannian parameters~\cite{K98}. However, in the Abelian case,
there exists an additional gauge invariant term, the so-called
Fayet-Iliopoulos or the D-term~\cite{FI}
\begin{equation}
{\cal L}_{F.I.}=\xi D=\int d^4\theta \xi V,  \label{FI}
\end{equation}
which requires special consideration. In Ref.~\cite{FNPRS}, it has
been shown that in the unbroken theory this term is not renormalized
provided the sum of hypercharges and their cubes equals zero. These
requirements guarantee the absence of chiral and gravity anomalies
and are usually satisfied in realistic models.

In case of a softly broken Abelian SUSY gauge theory, the F-I term
happens to be renormalized even if anomalies are cancelled. The RG
equation for $\xi$ depends not only on itself, but on the other
soft breaking parameters (the soft mass of chiral scalars $m^2$,
the soft triple coupling $A^{ijk}$ and the gaugino masses $M_i$).
Recently, the renormalization  of $\xi$  has been performed up to
three loops~\cite{JJ00,JJP} using the component approach and/or
superfields with softly broken Feynman rules. Here, following our
main idea that renormalizations of a softly broken SUSY theory are
completely defined by a rigid one, we suggest the method to get
the renormalization of the F-I term directly from the unbroken
theory\footnote{We have used the program DIANA~\cite{T} for the
generation of Feynman diagrams and the package MINCER~\cite{LTV}
for the evaluation of three and four loop diagrams}.

\section{Renormalization of the Fayet-Iliopoulos Term}
Consider an arbitrary N=1 SUSY gauge theory with the rigid
Lagrangian
\begin{eqnarray}
{\cal L}_{rigid} &=& \int d^2\theta~\frac{1}{4g^2}{\rm
Tr}W^{\alpha}W_{\alpha} + \int d^2\bar{\theta}~\frac{1}{4g^2}{\rm
Tr} \bar{W}_{\dot \alpha}\bar{W}^{\dot \alpha} \label{rigidlag} \\
&+& \int d^2\theta d^2\bar{\theta} ~~\bar{\Phi}^{j}
(e^{V})^i_j\Phi_i + \int
 d^2\theta ~~{\cal W} + \int d^2\bar{\theta} ~~\bar{\cal W},  \nonumber
\end{eqnarray}
where
 $$W_{\alpha}=-\frac 14\bar D^2e^{-V}D_\alpha e^V,
\ \ \ \bar W_{\dot \alpha}=-\frac 14 D^2e^{-V}\bar D_{\dot \alpha}
e^V,$$ are the gauge field strength tensors and the superpotential
${\cal W}$ has the form
\begin{equation}
{\cal  W}=\frac{1}{6}y^{ijk}\Phi_i\Phi_{\! j}\Phi_k +\frac{1}{2}
M^{ij}\Phi_i\Phi_{\! j}.\label{rigidsuppot}
\end{equation}
Soft SUSY breaking terms can be written as
\begin{eqnarray}
-{\cal L}_{soft-breaking} &=&\left[ \frac{M}{2\,g^2}\lambda\lambda
+\frac 16 A^{ijk} \phi_i\phi_j\phi_k+ \frac 12 B^{ij}\phi_i\phi_j
+h.c.\right] +(m^2)^i_j\bar\phi^{\,j}\phi_i.\label{sofl}
\end{eqnarray}

In the case of the Abelian gauge group the addition of the F-I term leads
to the modification of the Lagrangian in components.  The relevant
part of the Lagrangian is as follows:
\begin{equation}
{\cal L} =\displaystyle{ \frac{1}{2\, g^2}}D^2 +  \xi D + D\bar
\phi^{\,j} {\cal Y}_j^i\phi_i - \bar \phi^{\,j}
(m^2)_j^i\phi_i+...
\end{equation}
where  ${\cal Y\,}^i_j$ is the hypercharge matrix of the chiral
supermultiplet, and $(m^2)^i_j$ is a soft scalar mass. After
eliminating the auxiliary field $D$ this becomes
\begin{equation}
{\cal L} = -\bar\phi^{\,j} (\wbar m^2)_j^i\phi_i -
\displaystyle{\frac{1}{2}}g^2(\bar \phi^{\,j}{\cal
Y}_j^i\phi_i)^2+..., \label{ldel}
\end{equation}
where
\begin{equation}
 (\wbar m^2)^i_j = (m^2)^i_j + g^2 \xi {\cal
Y}^i_j. \label{mbar}
\end{equation}
From eqs.(\ref{ldel}) and (\ref{mbar}) it follows that the F-I
term gives an additional contribution to the renormalization of
the soft scalar mass $(m^2)^i_j$
\begin{equation}
[\beta_{\wbar m^2}]^i_j = [\beta_{m^2}]^i_j + \beta_{g^2} \xi
{\cal Y}^i_j+g^2 \beta_\xi(m^2,...) {\cal
Y}^i_j=[\beta_{m^2}]^i_j+g^2 \beta_\xi(\wbar m^2,...) {\cal
Y}^i_j.
\end{equation}
The last equality follows from the fact that eq.(\ref{ldel}) does
not contain $\xi$ explicitly and, hence, $\xi$ should be dropped from all
the expressions.

 There are four different types of contributions to
the renormalization of the F-I term in a softly broken theory:
those proportional to $(m^2)^i_j$, $M \bar M$, $A^{ijk} {\bar
A}_{lmn}$ and $M {\bar A}_{lmn}$ ($\bar M A^{ijk}$). Consider them
separately.

\subsection{The contribution proportional to $(m^2)^i_j$}
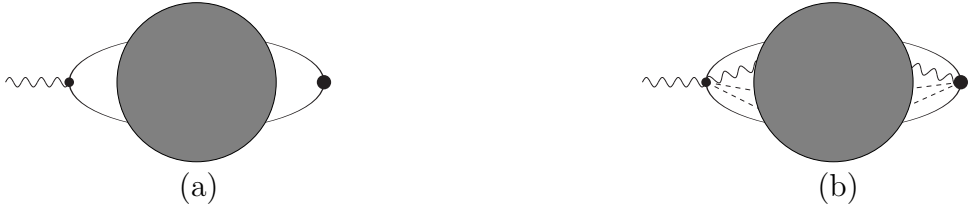
\begin{figure}[ht]
\begin{center} \begin{picture}(400,70)(0,0)
\SetScale{0.6} \Photon(20,50)(60,50){3}{4}
\Oval(140,50)(80,30)(90)
\GCirc(140,50){50}{0.5}
\Vertex(60,50){3} \Vertex(220,50){4.5}
\Photon(420,50)(460,50){3}{4} \Photon(460,50)(500,63){3}{4}
\Photon(620,50)(580,63){3}{4} \DashLine(460,50)(500,43){3}
\DashLine(460,50)(500,30){3} \DashLine(620,50)(580,45){3}
\DashLine(620,50)(580,32){3} \Oval(540,50)(80,30)(90)
\GCirc(540,50){50}{0.5} \Vertex(460,50){3} \Vertex(620,50){4.5}
\Text(85,-10)[c]{(a)} \Text(327,-10)[c]{(b)}
 \end{picture}
\end{center}
\caption{The diagrams giving a contribution to the F-I term renormalization
proportional to $(m^2)^i_j$ }\label{1}
\end{figure}

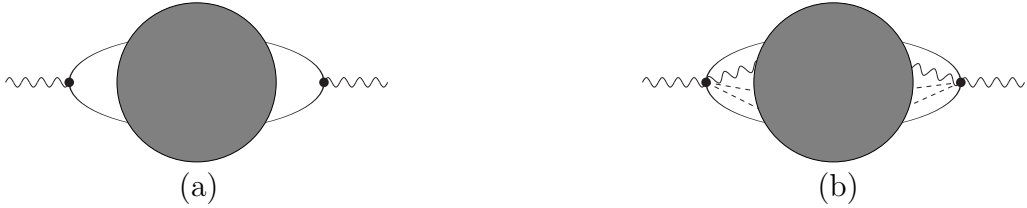
\begin{figure}[ht]
\begin{center} \begin{picture}(400,70)(0,0) \SetScale{0.6}
\Photon(20,50)(60,50){3}{4} \Oval(140,50)(80,30)(90)
\Photon(220,50)(260,50){3}{4} \GCirc(140,50){50}{0.5}
\Vertex(60,50){3} \Vertex(460,50){3} \Vertex(220,50){3}
\Photon(620,50)(660,50){3}{4} \Vertex(620,50){2.5}
\Vertex(620,50){3} \Photon(420,50)(460,50){3}{4}
\Photon(620,50)(580,63){3}{4} \Photon(460,50)(500,63){3}{4}
\DashLine(460,50)(500,43){3} \DashLine(460,50)(500,30){3}
\DashLine(620,50)(580,45){3} \DashLine(620,50)(580,32){3}
\Oval(540,50)(80,30)(90)
\GCirc(540,50){50}{0.5}
\Text(85,-10)[c]{(a)} \Text(327,-10)[c]{(b)}
\end{picture}   \end{center}
\caption{The diagrams giving a contribution to the Abelian vector
superfield
 renormalization}\label{2}
\end{figure}

We start with the contribution proportional to the soft scalar
mass. To find it in a superfield formalism, it is necessary to
calculate the diagrams shown in Fig.\ref{1}, where the dot on the
line in Fig.\ref{1}a means a softly broken superpropagator of a chiral
superfield and the vertex without external line means the vertex
proportional to  $(m^2)^i_j \theta^2 \bar \theta^2$.
Compare these diagrams with those giving contribution to the Abelian vector superfield
 renormalization. In terms of superfields, one has the diagrams shown in Fig.\ref{2}.

It is obvious that all the fields in a supermultiplet are
renormalized similarly. Consider the diagrams in components with
external lines being the D-components\footnote{Since we make our
calculations in superfields, in the corresponding component
diagrams one has to take into account all the fields from a vector
supermultiplet (see Appendix).}. In this case, the vertex in
diagram \ref{2}.a has the form
\begin{equation}
{\cal Y}^i_j D\bar \phi^{\,j} \phi_i=\int d^4\theta {\cal Y}^i_j
D\theta^2\bar \theta^2\bar \Phi^{\,j} \Phi_i.
\end{equation}
Inserting in this equation $(m^2)^i_j$ instead of ${\cal Y}^i_j
D$, one gets
\begin{equation}
\int d^4\theta {(m^2)}^i_j \theta^2\bar \theta^2\bar \Phi^{\,j}
\Phi_i= {(m^2)}^i_j \bar \phi^{\,j} \phi_i,
\end{equation}
which is nothing else but insertion of the soft scalar mass
from eq.(\ref{sofl}) into the scalar propagator. Hence, the
contribution of diagram \ref{2}.a to the renormalization of the
vector superfield is the same as the contribution of diagram
\ref{1}.a to that of $\xi$ from eq.(\ref{FI}) with the replacement of the hypercharge
${\cal Y}^i_j$ by the soft mass ${(m^2)}^i_j$.

Analogously, for diagram \ref{1}.b with a vertex with more than
one vector superfield, one can relate it to the diagram \ref{2}.b
\begin{eqnarray}
&&{\cal Y_\alpha}^i_k {\cal Y_\beta}^k_j D_\alpha\bar \phi^{\,j}
\phi_i C_\beta= \int d^4\theta {\cal Y_\alpha}^i_k {\cal
Y_\beta}^k_j D_\alpha \theta^2\bar \theta^2\bar\phi^{\,j} \phi_i
C_\beta= \int d^4\theta {\cal Y_\alpha}^i_k {\cal Y_\beta}^k_j
D_\alpha \theta^2\bar \theta^2\bar\Phi^j \Phi_i V_\beta
\Rightarrow\nonumber\\&& \int d^4\theta {(m^2)}^i_k{\cal
Y_\beta}^k_j\theta^2\bar \theta^2\bar \Phi^j \Phi_iV_\beta= \int
d^4\theta {(m^2)}^i_k {\cal Y_\beta}^k_j\theta^2\bar \theta^2\bar
\phi^{\,j} \phi_i C_\beta= {(m^2)}^i_k {\cal Y_\beta}^k_j\bar \phi^{\,j}
\phi_i C_\beta,
\end{eqnarray}
where $C_\beta$ stands for the lowest component of the vector
superfield $V_\beta$.
 \begin{figure}[ht]
\begin{center} \begin{picture}(200,80)(0,0)
\SetScale{0.6} \Photon(20,50)(60,50){3}{4}
\Oval(140,50)(80,30)(90)
\GCirc(140,50){50}{0.5}
\Photon(220,50)(260,50){3}{4} \Vertex(60,50){2.5} \Vertex(183,75){2.5}
\Vertex(183,25){2.5} \Vertex(220,50){2.5}
\LongArrowArcn(190,50)(40,70,30)
\LongArrowArcn(190,50)(40,330,290) \Text(116,53)[r]{1}
\Text(116,10)[r]{3} \Text(123,32)[l]{2} \Text(134,58)[l]{$p_1$}
\Text(135,9)[l]{$p_2$} \Text(150,40)[l]{$p$}
\end{picture}   \end{center}
\caption{An example of the Abelian vector superfield propagator diagram}\label{3}
\end{figure}
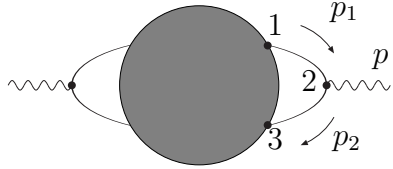

One can get the same rule of correspondence in a different way.
Consider the diagram shown in Fig.\ref{3}. Take the part of it
from the vertex 1 to the vertex 3
\begin{equation}
\int d^4(\theta_1 \theta_2 \theta_3) \frac{1}{p_1^2}\bar
D^2_{1,p_1} D^2_{2,-p_1}\delta_{12} {\cal Y}^i_j V(2,p)
\frac{1}{p_2^2}\bar D^2_{2,p_2} D^2_{3,-p_2}\delta_{23},
\end{equation}
and integrate by parts along the matter field propagator ($2\to
3$). This gives
\begin{equation}
\int d^4(\theta_1 \theta_2 \theta_3)\left( D^2_{2,p_2} \bar
D^2_{2,p_2} \frac{1}{p_1^2 p_2^2}{\cal Y}^i_j V(2,p) \bar
D^2_{1,p_1} D^2_{2,-p_1}\delta_{12}\right) \delta_{23}.
\end{equation}
Integration over $\theta_3$ and the substitution
$V(2,p)\rightarrow D \theta_2^2\bar \theta_2^2$ give ($p=0$,
$p_1=p_2$)
\begin{equation}
\int d^4(\theta_1 \theta_2) \frac{1}{p_1^2 p_1^2}D^2_{2,p_1} \bar
D^2_{2,p_1} {\cal Y}^i_j  D \theta_2^2\bar \theta_2^2
D^2_{2,-p_1}\bar D^2_{2,-p_1}\delta_{12},
\end{equation}
which coincides with the softly broken matter field superpropagator
with the substitution ${\cal Y}^i_j D\rightarrow
(m^2)^i_j$~\cite{HN}
$$\frac{m^2}{p^2(p^2-m^2)} D^2 \bar D^2  \theta^2\bar \theta^2
D^2 \bar D^2 \delta_{12}.$$

Hence, in a general case, one has to calculate the self-energy
diagrams for the vector superfield  and  in the resulting
expression to replace the hypercharge ${\cal Y}^i_j$,
corresponding to the external line, by the soft scalar mass
$(m^2)^i_j$.

 Using the results of Ref.\cite{JJN} and the above
formulated rule, after some algebraic manipulations and taking into
account the gauge invariance of the
Lagrangian~(\ref{rigidlag},\ref{rigidsuppot})
\begin{equation}
y^{ijn}{\cal Y}_n^k+y^{ink}{\cal Y}_n^j +y^{njk}{\cal Y}_n^i=0, \
\ \ \ \ \ (m^2)^i_k{\cal Y}_j^k={\cal Y}_k^i(m^2)^k_j,  \label{gi}
\end{equation}
one can quickly get the contribution of the soft mass $(m^2)^i_j$
to the renormalization of the F-I term, which coincides with
that from Ref.~\cite{JJP}.

\subsection{The contribution proportional to $A^{ijk}\bar A_{lmn}$}
There is another possibility of getting the desired result from the
unbroken theory, namely, to consider the propagator of the matter
superfield. It is known that, provided the anomalous dimension of the
matter field, one can get the beta function for the soft mass
$(m^2)^i_j$ acting on anomalous dimension by the differential
operator $D_2$~\cite{AKK,KV} (or $\cal O$~\cite{JJ}). The action
of this operator means that in the self energy diagram one has to
replace each  pair of  Yukawa couplings of opposite chirality
$y^{ijk}$ ($\bar y_{ijk}$) by the soft triple couplings
$A^{ijk}$ ($\bar A_{ijk}$) with the same indices (the term
proportional to $A^{ijk}$ ($\bar A_{ijk}$)), or to insert in each
line the soft mass term $(m^2)^i_j$ and contract the indices (the
term proportional to $(m^2)^i_j$). Consider how this procedure
works in components.

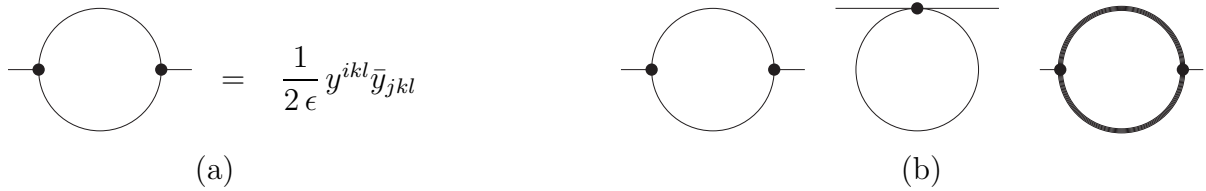
\begin{figure}[ht]\vspace{-0.5cm}
\begin{center} \begin{picture}(540,80)(0,0)
\SetScale{0.77} \Line(25,30)(40,30) \CArc(70,30)(30,0,360)
\Line(100,30)(115,30) \Vertex(40,30){3} \Vertex(100,30){3}
\Text(100,20)[l]{$=\;\;\;{\displaystyle \frac{1}{2\,\epsilon}}
\, y^{ikl}\bar y_{jkl}$} \Line(325,30)(340,30) \CArc(370,30)(30,0,360)
\Line(400,30)(415,30) \Vertex(340,30){3} \Vertex(400,30){3}
\Line(430,60)(510,60) \CArc(470,30)(30,0,360) \Vertex(470,60){3}
\Line(530,30)(540,30) \CArc(570,30)(30,0,360)
\CArc(570,30)(30.5,0,360)\CArc(570,30)(31,0,360)
\CArc(570,30)(29.5,0,360)\CArc(570,30)(29,0,360)
\Line(600,30)(610,30) \Vertex(540,30){3} \Vertex(600,30){3}
\Text(90,-15)[l]{(a)}
\Text(357,-15)[l]{(b)}
\end{picture}
\end{center}
 \caption{The one loop
matter superfield propagator diagram (a) and the corresponding
component diagrams (b) giving a contribution to ${[\beta_{m^2}]^i_j}^{(1)}$.
The bold line denotes the spinor fields}\label{4}
\end{figure}

In one loop order in the unbroken theory there is only one superfield
diagram shown in Fig.\ref{4}a.
In a softly broken theory, it leads to the following  beta function
for the soft mass:
\begin{equation}
\left[\beta_{m^2}\right]^{i\ (1)}_j = \frac{1}{2}A^{ikl}\bar
A_{jkl}+y^{inl}(m^2)^k_n\bar y_{jkl}
+\frac{1}{4}(m^2)^i_ny^{nkl}\bar y_{jkl}+\frac{1}{4}y^{ikl}\bar
y_{nkl}(m^2)^n_j.\label{rgm1}
\end{equation}
In components, the same result comes from the three diagrams
(Fig.\ref{4}b). The first two diagrams give the contribution
corresponding to the first and second terms in~(\ref{rgm1}),
respectively. However, the second diagram is the tadpole and this
is the same tadpole which gives the contribution to the
renormalization of the F-I term! One has only to make the
replacement $y^{inl} \bar y_{jkl}\rightarrow {\cal
Y}^n_k\delta^i_j$ in the second term of eq.(\ref{rgm1}) to get the
 contribution to the renormalization of the F-I term 
proportional to the soft mass $(m^2)^i_j$.
\footnote{ In one and two loops, there are only terms proportional 
to the soft scalar mass $(m^2)^i_j$ due to the requirement of  
anomaly cancellation. All the other contributions appear 
starting from the three loop level.}
\begin{figure}[h]\vspace{-0.5cm}
\begin{center} \begin{picture}(540,80)(0,0)
\SetScale{0.77} \Line(30,40)(42,40) \CArc(70,30)(30,0,360)
\Line(98,40)(110,40) \Line(42,20)(98,20) \Vertex(42,40){3}
\Vertex(98,40){3} \Vertex(42,20){3} \Vertex(98,20){3}
\Text(100,20)[l]{$=\;\;{\displaystyle \frac{1}{4\,\epsilon}}\, y^{ikm}\bar
y_{mst}y^{nst}\bar y_{jkn}$} \Line(330,60)(410,60)
\CArc(370,30)(30,0,360) \CArc(370,30)(30.5,180,360)
\CArc(370,30)(31,180,360) \CArc(370,30)(29.5,180,360)
\CArc(370,30)(29,180,360) \Line(340,30)(400,30)
\Line(340,30.5)(400,30.5) \Line(340,31)(400,31)
\Line(340,29.5)(400,29.5) \Line(340,29)(400,29) \Vertex(370,60){3}
\Vertex(340,30){3} \Vertex(400,30){3} \Line(430,40)(442,40)
\CArc(470,30)(30,0,360) \CArc(470,30)(30.5,0,360)
\CArc(470,30)(31,0,360) \CArc(470,30)(29.5,0,360)
\CArc(470,30)(29,0,360) \Line(498,40)(510,40)
\Line(442,20)(498,20) \Vertex(442,40){3} \Vertex(498,40){3}
\Vertex(442,20){3} \Vertex(498,20){3} \Line(530,40)(610,40)
\CArc(570,30)(30,0,360) \Vertex(542,40){3} \Vertex(598,40){3}
\Text(90,-15)[l]{(a)} \Text(357,-15)[l]{(b)}
\end{picture} \end{center}
 \caption{The two loop superfield diagram containing only Yukawa couplings (a)
 and the corresponding component diagrams (b) giving a contribution to ${[\beta_{m^2}]^i_j}^{(2)}$
}\label{6}
\end{figure}
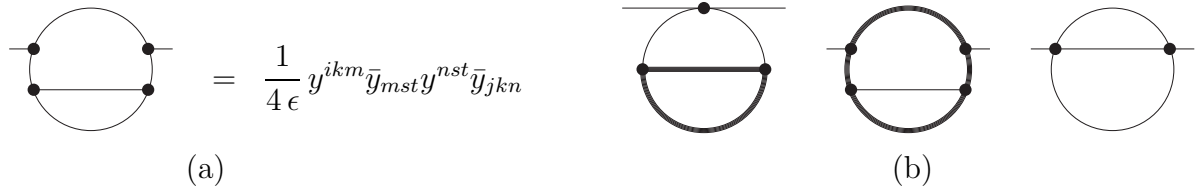
Analogously, in two loops a single superfield diagram containing
only the Yukawa couplings is shown in Fig.\ref{6}a, and the
corresponding soft beta function reads (the contribution
proportional to $(m^2)^i_j$)
\begin{eqnarray}
\left[\beta_{m^2}\right]^{i\ (2)}_j &=& \frac12
y^{ikl}(m^2)^m_l\bar y_{mst}y^{nst}\bar y_{jkn} +\frac12
y^{ikm}\bar y_{mst}y^{nst}(m^2)^l_n\bar y_{jkl}
+ y^{ikm}\bar y_{mlt}(m^2)^l_sy^{nst}\bar y_{jkn}\nonumber\\
&\hspace*{-2cm}+&\hspace*{-1cm}\frac12 (m^2)^k_l y^{ilm}\bar
y_{mst}y^{nst}\bar y_{jkn} +\frac14 (m^2)^i_l y^{lkm}\bar
y_{mst}y^{nst}\bar y_{jkn} +\frac14  y^{ikm}\bar
y_{mst}y^{nst}\bar y_{lkn}(m^2)^l_j. \label{rgm2}
\end{eqnarray}

Again in components, this result comes from the three diagrams in
Fig.\ref{6}b. The first two terms in~(\ref{rgm2}) come from the
first diagram and only from it. The rest diagrams give a
contribution to the other terms. Again, like in one loop order,
the first diagram is the same tadpole as in the D-term
renormalization. Besides, one can notice, that if being interested
in this analogy with the D-term renormalization, one may consider
only those superfield diagrams  where external lines are connected by a
single matter superfield line.

In three loops, one has two diagrams like that which shown in Fig.\ref{8}.
\begin{figure}[ht]
\begin{center} \begin{picture}(500,60)(0,0)
\SetScale{0.77} \Line(20,50)(48,50) \CArc(70,30)(30,0,360)
\Line(92,50)(120,50) \Line(40,30)(100,30) \Line(44,15)(96,15)
\Vertex(48,50){2.5} \Vertex(92,50){2.5} \Vertex(40,30){2.5}
\Vertex(100,30){2.5} \Vertex(44,15){2.5} \Vertex(96,15){2.5}
\Text(160,15)[c]{$={\displaystyle \frac{1}{3\,\epsilon}}\, y^{ikm} \bar
y_{mpr}y^{rst}\bar y_{qst}y^{npq}\bar y_{jkn}$}
\Line(320,50)(348,50) \CArc(370,30)(30,0,360)
\Line(392,50)(420,50) \Line(342,40)(362,2) \Line(398,40)(378,2)
\Vertex(348,50){2.5} \Vertex(392,50){2.5} \Vertex(342,40){2.5}
\Vertex(398,40){2.5} \Vertex(362,2){2.5} \Vertex(378,2){2.5}
\Text(399,15)[c]{$={-\displaystyle \frac{1}{12\,\epsilon}}\, y^{ikm} \bar
y_{mst}y^{lst}\bar y_{lpq}y^{npq}\bar y_{jkn}$}
\end{picture}
\end{center}
\caption{The three loop superfield diagrams containing only Yukawa
couplings}\label{8}
\end{figure}
 The contribution from these diagrams to the soft beta function is
(only the one proportional to $A^{ijk}\bar A_{lmn}$)
\begin{eqnarray}
\left[\beta_{m^2}\right]^{i\ (3)}_j &=& y^{ikm} \bar
A_{mpr}A^{rst}\bar y_{qst}y^{npq}\bar y_{jkn} +y^{ikm} \bar
y_{mpr}y^{rst}\bar A_{qst}A^{npq}\bar y_{jkn}
+y^{ikm} \bar A_{mpr}y^{rst}\bar y_{qst}A^{npq}\bar y_{jkn}\nonumber\\
&+&y^{ikm} \bar y_{mpr}A^{rst}\bar A_{qst}y^{npq}\bar
y_{jkn}+...
\nonumber\\
&-&\frac{1}{4}y^{ikm}\bar A_{mst}A^{lst}\bar y_{lpq}y^{npq}\bar
y_{jkn} -\frac{1}{4}y^{ikm}\bar y_{mst}y^{lst}\bar
A_{lpq}A^{npq}\bar y_{jkn}
-\frac{1}{4}y^{ikm}\bar A_{mst}y^{lst}\bar y_{lpq}A^{npq}\bar y_{jkn}\nonumber\\
&-&\frac{1}{4}y^{ikm}\bar y_{mst}A^{lst}\bar A_{lpq}y^{npq}\bar
y_{jkn}+... \label{rgm3}
\end{eqnarray}
The corresponding contributions in components come from the
diagrams shown in Fig.\ref{9}.
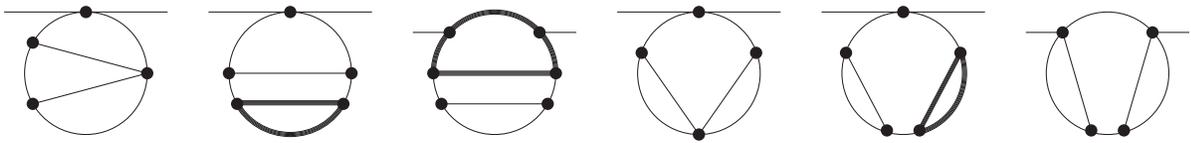
\begin{figure}[ht]\vspace{-0.5cm}
\begin{center} \begin{picture}(560,80)(0,0)
\SetScale{0.77}
\Line(30,60)(110,60) \CArc(70,30)(30,0,360) \Line(100,30)(44,45)
\Line(100,30)(44,15) \Vertex(100,30){3} \Vertex(70,60){3}
\Vertex(44,15){3} \Vertex(44,45){3}
\Line(130,60)(210,60)
\CArc(170,30)(30,0,360)
\CArc(170,30)(30.5,207,333)\CArc(170,30)(29.5,207,333)
\CArc(170,30)(31,207,333)\CArc(170,30)(29,207,333)
\Line(140,30)(200,30)
\Line(144,15.5)(196,15.5)
\Line(144,15)(196,15) \Line(144,16)(196,16)
\Line(144,14.5)(196,14.5) \Line(144,16.5)(196,16.5)
\Vertex(170,60){3} \Vertex(140,30){3}
\Vertex(200,30){3} \Vertex(144,15){3} \Vertex(196,15){3}
\Line(230,50)(248,50) \CArc(270,30)(30,0,360)
\CArc(270,30)(30.5,0,180) \CArc(270,30)(31,0,180)
\CArc(270,30)(29.5,0,180) \CArc(270,30)(29,0,180)
\Line(292,50)(310,50) \Line(240,30)(300,30)
\Line(240,29.5)(300,29.5) \Line(240,29)(300,29)
\Line(240,30.5)(300,30.5) \Line(240,31)(300,31)
\Line(244,15)(296,15) \Vertex(248,50){3} \Vertex(292,50){3}
\Vertex(240,30){3} \Vertex(300,30){3} \Vertex(244,15){3}
\Vertex(296,15){3}
\Line(330,60)(410,60) \CArc(370,30)(30,0,360) \Line(342,40)(370,0)
\Line(398,40)(370,0) \Vertex(370,60){3} \Vertex(342,40){3}
\Vertex(398,40){3} \Vertex(370,0){3}
\Line(430,60)(510,60)
\CArc(470,30)(30,0,360)
\CArc(470,30)(30.5,285,20)\CArc(470,30)(31,285,20)
\CArc(470,30)(29.5,285,20)\CArc(470,30)(29,285,20)
\Line(442,40)(462,2) \Line(498,40)(478,2)
\Line(498.5,40)(478.5,2) \Line(497.5,40)(477.5,2)
\Line(499,40)(479,2) \Line(497,40)(477,2)
\Vertex(470,60){3} \Vertex(442,40){3} \Vertex(498,40){3}
\Vertex(462,2){3} \Vertex(478,2){3}
\Line(530,50)(548,50)
\CArc(570,30)(30,0,360) \Line(592,50)(610,50) \Line(548,50)(562,2)
\Line(592,50)(578,2) \Vertex(548,50){3} \Vertex(592,50){3}
\Vertex(562,2){3} \Vertex(578,2){3}
\end{picture} \end{center}
\caption{The component diagrams corresponding to Fig.\ref{8} and
contributing to ${[\beta_{m^2}]^i_j}^{(3)}$}\label{9}
\end{figure}
Again, it is easy to see that the first and the third lines
in~(\ref{rgm3}), when calculated in components, come from the
tadpole diagrams while all the rest from the other diagrams.
Taking these lines and performing the replacement
$$y^{ikm}\bar y_{jkn}\rightarrow {\cal Y}^m_n\delta^i_j,$$
after making use of~(\ref{gi}), valid also for replacement of all
the Yukawa vertices $y^{ijk}$ by the corresponding soft triple
couplings $A^{ijk}$, one obtains the contribution to the
renormalization of the F-I term proportional to $A\bar A y\bar y$
coinciding with that of Ref.~\cite{JJP}.

\begin{figure}[ht]
\begin{center} \begin{picture}(460,50)(0,0)
\SetScale{0.77}
\Line(30,50)(48,50) \CArc(70,30)(30,0,360)
\Line(92,50)(110,50) \Photon(40,30)(100,30){3}{4.5}
\Line(44,15)(96,15) \Vertex(48,50){2.5} \Vertex(92,50){2.5}
\Vertex(40,30){2.5} \Vertex(100,30){2.5} \Vertex(44,15){2.5}
\Vertex(96,15){2.5}

\Line(130,50)(148,50) \CArc(170,30)(30,0,360)
\Line(192,50)(210,50) \Line(140,30)(200,30)
\Photon(144,15)(196,15){3}{4.5} \Vertex(148,50){2.5}
\Vertex(192,50){2.5} \Vertex(140,30){2.5} \Vertex(200,30){2.5}
\Vertex(144,15){2.5} \Vertex(196,15){2.5}

\Line(230,50)(248,50) \CArc(270,30)(30,0,360)
\Line(292,50)(310,50) \Photon(242,40)(262,2){3}{4.5}
\Line(298,40)(278,2) \Vertex(248,50){2.5} \Vertex(292,50){2.5}
\Vertex(242,40){2.5} \Vertex(298,40){2.5} \Vertex(262,2){2.5}
\Vertex(278,2){2.5}

\Line(330,50)(348,50) \CArc(370,30)(30,0,360)
\Line(392,50)(410,50) \Line(340,30)(400,30)
\Photon(370,30)(370,0){3}{2.5} \Vertex(348,50){2.5}
\Vertex(392,50){2.5} \Vertex(340,30){2.5} \Vertex(400,30){2.5}
\Vertex(370,30){2.5} \Vertex(370,0){2.5}

\Line(430,50)(448,50) \CArc(470,30)(30,0,360)
\Line(492,50)(510,50) \Line(440,30)(470,30)
\Photon(470,30)(500,30){3}{2.5} \Line(470,30)(470,0)
\Vertex(448,50){2.5} \Vertex(492,50){2.5} \Vertex(440,30){2.5}
\Vertex(500,30){2.5} \Vertex(470,30){2.5} \Vertex(470,0){2.5}
\end{picture}  \end{center}
\caption{The three loop superfield diagrams with one internal
vector line}\label{10}
\end{figure}
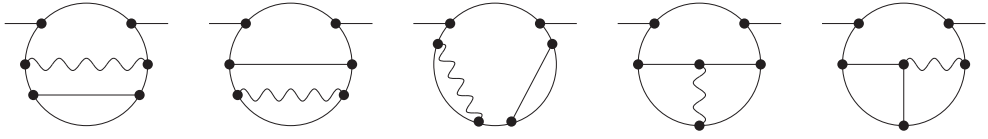

Unfortunately, this procedure does not work always. To see this,
consider three loop diagrams, contributing to the matter superfield
propagator, with one internal vector line as shown in
Fig.\ref{10}.

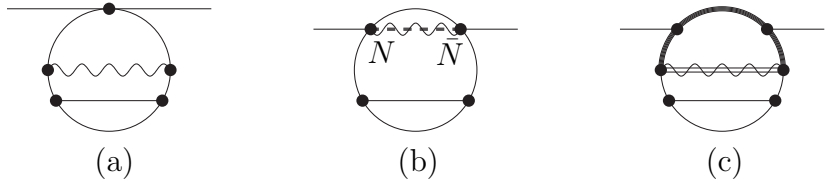
\begin{figure}[ht]\vspace{-0.5cm}
\begin{center} \begin{picture}(380,80)(0,0)
\SetScale{0.77}
\Line(20,60)(120,60) \CArc(70,30)(30,0,360)
\Photon(40,30)(100,30){3}{4.5} \Line(44,15)(96,15)
\Vertex(70,60){3} \Vertex(40,30){3} \Vertex(100,30){3}
\Vertex(44,15){3} \Vertex(96,15){3}\Line(170,50)(198,50)
\CArc(220,30)(30,0,360) \Line(242,50)(270,50) \Text(56,-12)[c]{(a)}
\DashLine(242,50)(198,50){4}
\DashLine(242,49.5)(198,49.5){4}
\DashLine(242,50.5)(198,50.5){4}
\Photon(242,50)(198,50){3}{3.5} \Line(194,15)(246,15)
\Vertex(198,50){3} \Vertex(242,50){3} \Vertex(194,15){3}
\Vertex(246,15){3} \Text(157,30)[]{$N$} \Text(183,30)[]{$\bar N$}
\Line(320,50)(348,50) \CArc(370,30)(30,0,360)
\CArc(370,30)(30.5,0,180) \CArc(370,30)(29.5,0,180)
\CArc(370,30)(31,0,180) \CArc(370,30)(29,0,180)
\Line(392,50)(420,50) \Line(340,31)(400,31) \Line(340,29)(400,29)
\Text(171,-12)[c]{(b)} \Photon(340,30)(400,30){3}{4.5}
\Line(344,15)(396,15) \Vertex(348,50){3} \Vertex(392,50){3}
\Vertex(340,30){3} \Vertex(400,30){3} \Vertex(344,15){3}
\Vertex(396,15){3} \Text(287,-12)[c]{(c)}
\end{picture}
 \end{center}
 \caption{ The component diagrams corresponding to the first diagram in
 Fig.\ref{10}. The dashed line denotes an auxiliary scalar field
 $N$ from a vector supermultiplet and the double line denotes a
 gaugino field}
\label{11}
\end{figure}

\noindent Consider  the first diagram. The simple pole is equal to
\begin{equation}
-4 g^2\frac{1}{3\epsilon}y^{ikl}{\cal Y}^m_l \bar y_{mst}
y^{nst}{\cal Y}^p_n\bar y_{jkp}\label{d1aa1}
\end{equation}
and gives the following contribution to the soft scalar mass beta
function (only the term proportional to $A\bar A$)
\begin{equation}
\left[\beta_{m^2}\right]^{i\ (3)}_j =-4 g^2y^{ikl}{\cal Y}^m_l
\bar A_{mst} A^{nst}{\cal Y}^p_n\bar y_{jkp}-4 g^2A^{ikl}{\cal
Y}^m_l \bar A_{mst} y^{nst}{\cal Y}^p_n\bar
y_{jkp}+...\label{rgm3aa1}
\end{equation}
In components one has three different diagrams shown in
Fig.\ref{11}.
All three diagrams give a contribution to the first term
in~(\ref{rgm3aa1}).
So one cannot directly extract
 the contribution from the tadpole graphs. However, to get the
renormalization of the D-term proportional to $g^2A\bar A$ it is
sufficient to calculate only two diagrams, namely \ref{11}.b and
\ref{11}.c. In the rest of the diagrams in Fig.\ref{10} (except the
first) the contribution from the tadpole graphs can be figured out
from the superfield diagrams. The simple poles for diagrams
\ref{11}.b and \ref{11}.c omitting the tensor structure are,
respectively,
$$9.b= \frac{2}{3\,\epsilon},\ \ \ \ \ \ \ \
9.c=\frac{2}{\epsilon}.$$
 Subtracting these expressions from~(\ref{d1aa1}) one gets the result
for the diagram \ref{11}.a
$$
-4 g^2\frac{1}{\epsilon}y^{ikl}{\cal Y}^m_l \bar y_{mst}
y^{nst}{\cal Y}^p_n\bar y_{jkp},
$$
which, after the replacement $y^{ikl}\bar
y_{jkp}\rightarrow {\cal Y}^l_p\delta^i_j,$ gives the contribution
to the beta function of $\xi$ equal to $-12g^2{\cal Y}^l_p{\cal
Y}^m_l \bar A_{mst} A^{nst}{\cal Y}^p_n$. This term
 together with the results for the tadpole graphs obtained from the
superfield diagrams of Fig.\ref{10}, after reducing to the same
tensor structure gives the renormalization of the F-I term
proportional to $g^2A\bar A$ coinciding with that of
Ref.\cite{JJP}.

The same way one can determine the contributions proportional to
 $A^{ijk}\bar M$ ($\bar A^{ijk}M$).  As an
illustration of efficiency of this method, we present below the
calculation of the four loop contribution to the renormalization
of the F-I term proportional to $ A\bar A y \bar y y \bar y$.

\subsection{The contribution proportional to $A\,\bar M$ ($\bar A\,M$)}
This contribution to the F-I term renormalization can be
calculated in a way similar to the previous one. In this 
case, the analysis of the component
diagrams shows that one should consider the following four graphs
\begin{figure}[ht]
\begin{center} \begin{picture}(460,50)(0,0)
\SetScale{0.78} \Line(20,50)(48,50) \CArc(70,30)(30,0,360)
\Line(92,50)(120,50) \Photon(42,40)(62,2){3}{4.5}
\Line(98,40)(78,2) \Vertex(48,50){2.5} \Vertex(92,50){2.5}
\Vertex(42,40){2.5} \Vertex(98,40){2.5} \Vertex(62,2){2.5}
\Vertex(78,2){2.5} \Text(56,-12)[]{(a)}

\Line(170,50)(198,50) \CArc(220,30)(30,0,360)
\Line(242,50)(270,50) \Line(190,30)(250,30)
\Photon(194,15)(246,15){3}{4.5} \Vertex(198,50){2.5}
\Vertex(242,50){2.5} \Vertex(190,30){2.5} \Vertex(250,30){2.5}
\Vertex(194,15){2.5} \Vertex(246,15){2.5} \Text(174,-12)[]{(b)}

\Line(320,50)(348,50) \CArc(370,30)(30,0,360)
\Line(392,50)(420,50) \Line(340,30)(400,30)
\Photon(370,30)(370,0){3}{2.5} \Vertex(348,50){2.5}
\Vertex(392,50){2.5} \Vertex(340,30){2.5} \Vertex(400,30){2.5}
\Vertex(370,30){2.5} \Vertex(370,0){2.5} \Text(291,-12)[]{(c)}

\Line(470,50)(498,50) \CArc(520,30)(30,0,360)
\Line(542,50)(570,50) \Line(490,30)(520,30)
\Photon(520,30)(550,30){3}{2.5} \Line(520,30)(520,0)
\Vertex(498,50){2.5} \Vertex(542,50){2.5} \Vertex(490,30){2.5}
\Vertex(550,30){2.5} \Vertex(520,30){2.5} \Vertex(520,0){2.5}
\Text(408,-12)[]{(d)}
\end{picture}  \end{center}
\caption{Three loop superfield diagrams giving contribution proportional to
$A\,\bar M$}\label{12} \end{figure}
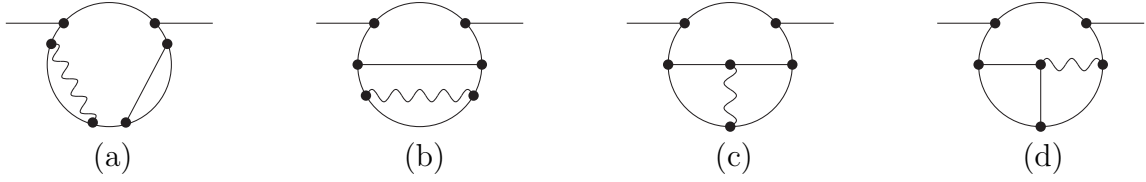 and in all
these diagrams the contribution of the tadpole graphs to the soft
scalar mass renormalization can be easily determined without any use of
the component calculations.

Notice, however, that in the Wess-Zumino gauge the first two
diagrams do not give any contribution. Still we have to take into
account all the fields from the vector supermultiplet. The results for
these diagrams are, respectively, (simple pole) :
\begin{center}
\begin{tabular}{|c|c|c|c|}\hline
 & & &  \\[-3mm]
a & b & c & d  \\[1mm]
 \hline
  & & &  \\
 $\;\;\displaystyle{-\frac{1}{6\epsilon}}\;\;$&
$\;\;-\displaystyle{\frac{2}{3\epsilon}}\;\;$&
$\;\;\displaystyle{\frac{1}{\epsilon}}\zeta(3)\;\;$&
$\;\;\displaystyle{\frac{2}{3\epsilon}}\;\;$\\
& & &  \\
\hline
\end{tabular}
\end{center}
which, after the replacement of the Yukawa vertices with an external lines
by a hypercharge in the proper part of the soft scalar mass
beta function and the reduction of tensor structures, give
the answer coinciding with that of Ref.~\cite{JJP}.

\subsection{The contribution proportional to $M\,\bar M$}
\begin{figure}[ht]\vspace{-0.8cm}
\begin{center} \begin{picture}(400,80)(0,0)
\SetScale{0.6} \Photon(420,50)(470,50){3}{4}
\CArc(505,50)(35,90,270)
\GCirc(540,50){50}{0.5}
\PhotonArc(575,50)(35,270,90){3}{8}
\Photon(660,50)(610,50){3}{4.5} \Vertex(470,50){3}
\Vertex(610,50){3}
\Text(380,55)[c]{$\bar D^2 D_\alpha$}
\Text(375,15)[c]{$D^\alpha$}
\Photon(20,50)(70,50){3}{4}
\CArc(105,50)(35,90,270)
\PhotonArc(175,50)(35,270,90){3}{8} \GCirc(140,50){50}{0.5}
\Vertex(70,50){3} \Vertex(210,50){6} \Text(87,-10)[c]{(a)}
\Text(330,-10)[c]{(b)}
\end{picture}
\end{center}
\caption{The tadpole graph (a) giving contribution proportional to $M\,\bar M$
and the corresponding self-energy vector superfield diagram (b)
}\label{13}
\end{figure}
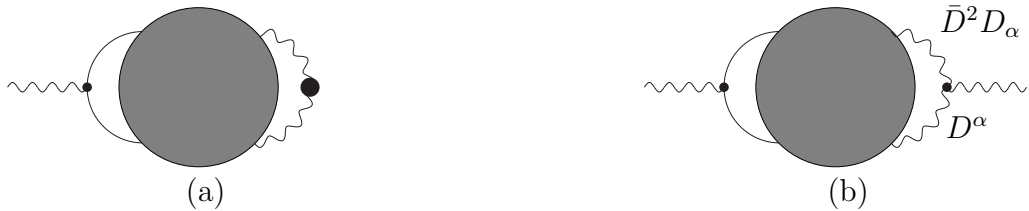

The contribution to the D-term renormalization proportional to
$M\,\bar M$ may come either from one of the vector lines ($M\,\bar
M$), or from two different lines ($M$ and $\bar M$). In the first
case, to get the result, one has to calculate in a rigid theory the
superfield diagrams shown in Fig.\ref{13}b, where in the triple
vector vertex the external line does not contain supercovariant
derivatives. This diagram corresponds to the tadpole graph
\ref{13}.a with a softly broken vector superpropagator proportional to
$M\,\bar M$
$$M\bar M \frac{D^{\alpha}\bar D^2\theta^2 \bar \theta^2 D^2 \bar
D^{\dot \beta} p_{\alpha \dot \beta}}{p^4(p^2-M \bar M)},$$
 and gives the same result as diagram Fig.\ref{13}a after
the replacement ${\cal Y}^i_j \to M\bar M \delta^i_j$ for the
hypercharge corresponding to the external vertex.

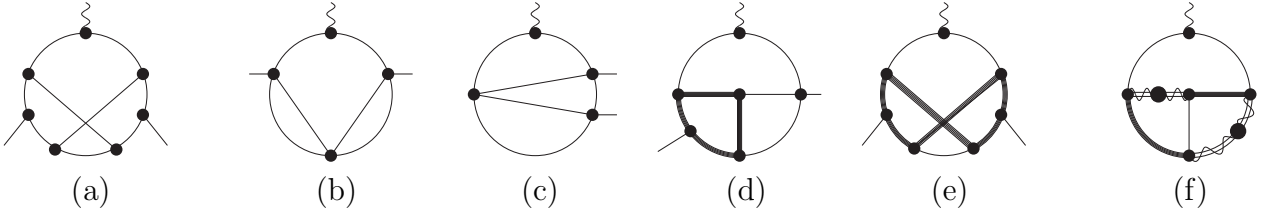
\begin{figure}[ht]
\begin{center} \begin{picture}(460,80)(0,0)
\SetScale{0.77} \Photon(30,65)(30,80){2}{2} \CArc(30,35)(30,0,360)
\Line(2,45)(45,8) \Line(58,45)(15,8) \Line(2,25)(-10,10)
\Line(58,25)(70,10) \Vertex(30,65){3} \Vertex(2,25){3}
\Vertex(58,25){3} \Vertex(58,45){3} \Vertex(2,45){3}
\Vertex(45,8){3} \Vertex(15,8){3} \Text(25,-10)[c]{(a)}

\Photon(150,65)(150,80){2}{2} \CArc(150,35)(30,0,360)
\Line(122,45)(150,5) \Line(178,45)(150,5) \Line(122,45)(110,45)
\Line(178,45)(190,45) \Vertex(150,65){3} \Vertex(178,45){3}
\Vertex(122,45){3} \Vertex(150,5){3} \Text(118,-10)[c]{(b)}

\Photon(250,65)(250,80){2}{2} \CArc(250,35)(30,0,360)
\Line(220,35)(278,25) \Line(278,45)(220,35) \Line(278,45)(290,45)
\Line(278,25)(290,25) \Vertex(250,65){3} \Vertex(220,35){3}
\Vertex(278,25){3} \Vertex(278,45){3} \Text(195,-10)[c]{(c)}

\Photon(350,65)(350,80){2}{2} \CArc(350,35)(30,0,360)
\CArc(350,35)(30.5,180,270) \CArc(350,35)(29.5,180,270)
\CArc(350,35)(31,180,270) \CArc(350,35)(29,180,270)
\Line(320,35)(350,35)
\Line(320,35.5)(350,35.5) \Line(320,34.5)(350,34.5)
\Line(320,36)(350,36) \Line(320,34)(350,34)
\Line(380,35)(350,35)
\Line(326,17)(310,7) \Line(350,35)(350,5) \Line(380,35)(390,35)
\Line(350.5,35)(350.5,5) \Line(349.5,35)(349.5,5)
\Line(351,35)(351,5) \Line(349,35)(349,5)
 \Vertex(350,65){3}
\Vertex(320,35){3} \Vertex(350,35){3} \Vertex(326,17){3}
\Vertex(350,5){3} \Vertex(380,35){3} \Text(273,-10)[c]{(d)}
 
\Photon(450,65)(450,80){2}{2} \CArc(450,35)(30,0,360)
\CArc(450,35)(30.5,158,240) \CArc(450,35)(29.5,158,240)
\CArc(450,35)(31,158,240) \CArc(450,35)(29,158,240)
\CArc(450,35)(30.5,300,22) \CArc(450,35)(29.5,300,22)
\CArc(450,35)(31,300,22) \CArc(450,35)(29,300,22)
\Line(422,45)(465,8) \Line(423,45)(466,8) \Line(421,45)(464,8)
\Line(424,45)(467,8) \Line(420,45)(463,8)
\Line(478,45)(435,8) \Line(479,45)(436,8) \Line(477,45)(434,8)
\Line(480,45)(434,8) \Line(476,45)(433,8)
\Line(422,25)(410,10) \Line(478,25)(490,10) \Vertex(450,65){3}
\Vertex(422,25){3} \Vertex(478,25){3} \Vertex(478,45){3}
\Vertex(422,45){3} \Vertex(464.5,8.5){3} \Vertex(435.5,8.5){3}
\Text(350,-10)[c]{(e)}

\Photon(570,65)(570,80){2}{2} \CArc(570,35)(30,0,270)
\CArc(570,35)(30.5,180,270) \CArc(570,35)(29.5,180,270)
\CArc(570,35)(31,180,270) \CArc(570,35)(29,180,270)
\CArc(570,35)(31,270,360) \CArc(570,35)(29,270,360)
\PhotonArc(570,35)(30,270,360){3}{4} \Photon(540,35)(570,35){3}{4}
\Line(540,36)(570,36) \Line(540,34)(570,34)
\Line(570,35.5)(600,35.5) \Line(570,34.5)(600,34.5)
\Line(570,36)(600,36) \Line(570,34)(600,34)
\Line(600,35)(570,35) \Line(570,35)(570,5) \Vertex(570,65){3}
\Vertex(540,35){3} \Vertex(570,35){3} \Vertex(555,35){4}
\Vertex(594,17){4} \Vertex(570,5){3} \Vertex(600,35){3}
\Text(441,-10)[c]{(f)}
\end{picture}\end{center}
\caption{The three loop vertex diagrams: (a) is the superfield
diagram, (b-e) are the corresponding component diagrams and (f) is
the needed tadpole diagram}\label{14}
\end{figure}

In the second case, one can proceed as follows. Consider the
superfield diagram of the vertex type (the interaction of a vector
superfield with matter) shown in Fig.\ref{14}a.

Rewriting the superfield diagram \ref{14}.a in components, one
finds four types of diagrams: \ref{14}.b and \ref{14}.c 
when cutting the external legs are the same component
diagrams which give a contribution to the renormalization of the
D-term proportional to $A\bar A y \bar y$; analogously, \ref{14}.d
gives the contribution proportional to $ g^2 \bar y A \bar M$ (with the
replacement of the fermion matter field propagator by that of
gaugino) and \ref{14}.e is analogous to \ref{14}.f, which we are
looking for, but with the insertion of mass into the gaugino lines.
Thus, subtracting from the superfield result \ref{14}.a the
already known component expressions \ref{14}.b-d, we get the
answer for diagram \ref{14}.e which, after the replacement of the
external Yukawa vertices by the gaugino mass, gives the desired
result for diagram \ref{14}.f.

One may have an impression that the calculations in components are
simpler. However, first of all, in our approach one practically does
not have to calculate anything new and, second, we just want to
emphasize that all the information on the renormalizations in a
softly broken theory can be extracted from the rigid one, though
for the case of the F-I term in a rather tricky way.

\section{Calculations in Four Loops}

\subsection{The contribution proportional to $(m^2)^i_j$}
To calculate the contribution proportional to $ (m^2) y \bar y y
\bar y y\bar y$, we follow the recipe of Sect.2.1 and use the
results of Ref.\cite{JJN}. After reduction of tensor structures
one finds
\begin{eqnarray}
\Delta\beta_{\xi}^{(4)}&=&-F^{(m^2)}_1-\frac{13}{2}F^{(m^2)}_2-\frac{10}{3}F^{(m^2)}_3
-\frac{5}{24}F^{(m^2)}_4+\frac {1}{2} \left(F^{(m^2)}_5+{\tilde
F}^{(m^2)}_5 \right) +\frac {3}{2}\left(F^{(m^2)}_6+{\tilde
F}^{(m^2)}_6 \right)\nonumber\\&+& \left(\frac {7}{6} - 2 \zeta(3)
\right)F^{(m^2)}_7+ \left(1 - 2 \zeta(3)\right)F^{(m^2)}_8-
\left(\frac {1}{2} + \zeta(3) \right)F^{(m^2)}_9
 -3 \zeta(3) F^{(m^2)}_{10},
\end{eqnarray}
where
$$ F^{(m^2)}_1={y}^{jkl}{\cal Y}_{j}^{i}{\bar y}_{ikm}y^{mnp}\bar y_{lnq}y^{qst}(m^2)_s^r\bar
y_{prt}\ \ \ \ \
F^{(m^2)}_2={y}^{jkl}{\cal Y}_{j}^{r}(m^2)_r^i{\bar y}_{ikm}y^{mnp}\bar y_{lnq}y^{qst}\bar y_{pst}$$
$$ F^{(m^2)}_3={y}^{jkl}{\cal Y}_{j}^{i}(m^2)_k^r{\bar y}_{irm}y^{mnp}\bar y_{lnq}y^{qst}\bar y_{pst}\ \ \ \ \
F^{(m^2)}_4={y}^{ikl}{\bar y}_{ikm}y^{mnp}{\cal Y}_{n}^{j}(m^2)^r_j\bar y_{lrq}y^{qst}\bar y_{pst}$$
$$ F^{(m^2)}_5={y}^{ikl}{\bar y}_{ikm}{\cal Y}^{m}_{j}y^{jnp}(m^2)_p^r\bar y_{lnq}y^{qst}\bar y_{rst}\ \ \ \ \
{\tilde F}^{(m^2)}_5={y}^{ikl}{\bar y}_{ikm}(m^2)^m_jy^{jnp}{\cal Y}_{p}^{r}\bar y_{lnq}y^{qst}\bar y_{rst}$$
$$ F^{(m^2)}_6={y}^{ikl}{\bar y}_{ikm}{\cal Y}^{m}_{j}(m^2)^j_ry^{rnp}\bar y_{lnq}y^{qst}\bar y_{pst}\ \ \ \ \
{\tilde F}^{(m^2)}_6={y}^{ikl}{\bar y}_{ikm}y^{mnp}\bar y_{lnq}y^{qst}{\cal Y}_{p}^{j}(m^2)_j^r\bar y_{rst}$$
$$ F^{(m^2)}_7={y}^{jkl}{\cal Y}_{j}^{i}(m^2)_k^r{\bar y}_{irm}y^{mpq}\bar y_{npq}y^{nst}\bar y_{lst}\ \ \ \ \
F^{(m^2)}_8={y}^{jkl}{\cal Y}_{j}^{r}(m^2)_r^i{\bar y}_{ikm}y^{mpq}\bar y_{npq}y^{nst}\bar y_{lst}$$
$$ F^{(m^2)}_9={y}^{ikl}{\bar y}_{ikm}{\cal Y}^{m}_{j}(m^2)^j_ry^{rpq}\bar y_{npq}y^{nst}\bar y_{lst}\ \ \ \ \
F^{(m^2)}_{10}={y}^{jln}{\cal Y}_{j}^{r}(m^2)_r^i{\bar y}_{ikm}y^{kps}\bar y_{lpq}y^{mqt}\bar y_{nst}$$

\subsection{The contribution proportional to $A\bar A$}
To calculate the contribution proportional to $A\bar A y \bar y y
\bar y$, according to our method, one has first of all to take the
results of the four loop calculation of the self-energy diagrams for the
matter superfield where the external lines are connected by a single
propagator of the matter superfield. There are six diagrams of this
sort.
\begin{figure}[ht]
\begin{center} \begin{picture}(500,50)(0,0)
\SetScale{0.77}
\Line(30,50)(48,50) \CArc(70,30)(30,0,360) \CArc(70,10)(30,20,160)
\Line(92,50)(110,50) \Line(50,34)(90,4) \Line(90,34)(50,4)
\Vertex(48,50){2.5} \Vertex(92,50){2.5} \Vertex(42,19){2.5}
\Vertex(98,19){2.5} \Vertex(51,33){2.5} \Vertex(88,6){2.5}
\Vertex(89,33){2.5} \Vertex(52,6){2.5} \Text(55,-12)[c]{(a)}

\Line(130,50)(148,50) \CArc(170,30)(30,0,360)
\Line(192,50)(210,50) \Line(141,36)(199,36) \Line(140,23)(200,23)
\Line(148,10)(192,10) \Vertex(148,50){2.5} \Vertex(192,50){2.5}
\Vertex(141,36){2.5} \Vertex(199,36){2.5} \Vertex(140.5,23){2.5}
\Vertex(199.5,23){2.5} \Vertex(148,10){2.5} \Vertex(192,10){2.5}
\Text(132,-12)[c]{(b)}

\Line(230,50)(248,50) \CArc(270,30)(30,0,360)
\Line(292,50)(310,50) \Line(241,36)(299,36)
\CArc(233,-6)(30,10,80) \CArc(307,-6)(30,100,170)
\Vertex(248,50){2.5} \Vertex(292,50){2.5} \Vertex(241,36){2.5}
\Vertex(299,36){2.5} \Vertex(240.5,23){2.5} \Vertex(299.5,23){2.5}
\Vertex(248,50){2.5} \Vertex(292,50){2.5} \Vertex(262,2){2.5}
\Vertex(278,2){2.5} \Text(209,-12)[c]{(c)}
\Line(330,50)(348,50) \CArc(370,30)(30,0,360) \CArc(370,10)(30,20,160)
\CArc(370,56)(30,230,310) \CArc(370,-16)(30,50,130)
\Line(392,50)(410,50) \Vertex(348,50){2.5} \Vertex(392,50){2.5}
\Vertex(342,19){2.5} \Vertex(398,19){2.5} \Vertex(351,33){2.5}
\Vertex(388,6){2.5} \Vertex(389,33){2.5} \Vertex(352,6){2.5}
\Text(287,-12)[c]{(d)}

\Line(430,50)(448,50) \CArc(470,30)(30,0,360)
\Line(492,50)(510,50) \CArc(424,12)(30,-12,60)
\CArc(507,-6)(30,100,170) \CArc(497,3)(30,86,185)
\Vertex(448,50){2.5} \Vertex(492,50){2.5} \Vertex(441,36){2.5}
\Vertex(500,33){2.5} \Vertex(453,5){2.5} \Vertex(499.5,23){2.5}
\Vertex(466,1){2.5} \Vertex(478,2){2.5} \Text(364,-12)[c]{(e)}

\Line(530,50)(548,50) \CArc(570,30)(30,0,360)
\CArc(518,25)(30,-25,35) \CArc(622,25)(30,145,205)
\CArc(570,-20)(30,53,127) \Line(592,50)(610,50) \Vertex(548,50){2.5}
\Vertex(592,50){2.5} \Vertex(545,14){2.5} \Vertex(595,14){2.5}
\Vertex(543,41){2.5} \Vertex(597,41){2.5} \Vertex(552,5){2.5}
\Vertex(588,5){2.5} \Text(441,-12)[c]{(f)}
\end{picture}   \end{center}
 \caption{The four loop self-energy superfield diagrams}\label{15}
\end{figure}
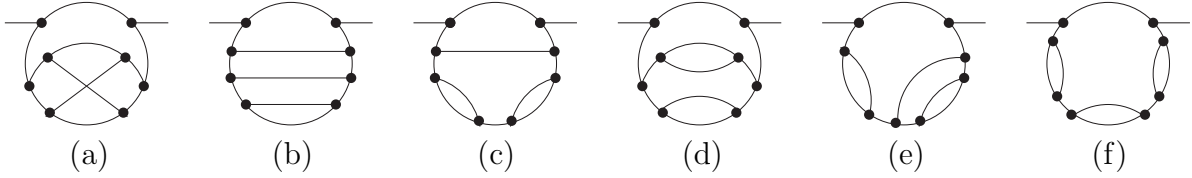
The results of their calculations are (single pole)
\begin{center}
\begin{tabular}{|c|c|c|c|c|c|}\hline
 & & & & &  \\[-3mm]
a & b & c & d & e & f \\[1mm]
\hline
  & & & & &  \\
 $\displaystyle{\frac{3}{8\epsilon}}\zeta(3)+\displaystyle{\frac{3}{16\epsilon}}\zeta(4)$&
$\;\;\displaystyle{\frac{5}{8\epsilon}}\;\;$&
$\displaystyle{-\frac{5}{48\epsilon}}+\displaystyle{\frac{1}{8\epsilon}}\zeta(3)$&
$-\displaystyle{\frac{5}{96\epsilon}}\;$&$
-\displaystyle{\frac{1}{6\epsilon}}\;$&
$\displaystyle{\frac{1}{32\epsilon}}
-\displaystyle{\frac{1}{16\epsilon}}\zeta(3) $\\
& & & & &  \\
\hline
\end{tabular}
\end{center}
The sum coincides with the proper part of Ref.\cite{AGKL}.

Acting on this sum by an operator $A^{ijk}\bar A_{lmn}
\displaystyle{\frac {\partial}{\partial y^{ijk}}}
\displaystyle{\frac {\partial}{\partial \bar y_{lmn}}}$, one
obtains the part of the beta function for the soft mass
$(m^2)^i_j$ proportional to $A\bar A y \bar y y \bar y y \bar y$.

Comparing the results with the corresponding component diagrams
containing two soft vertices of opposite chirality ($A^{ijk}$ and
$\bar A_{ijk}$), one finds that to extract the contribution of the
tadpole diagrams one need to calculate three diagrams
 \vspace{0.5cm}

 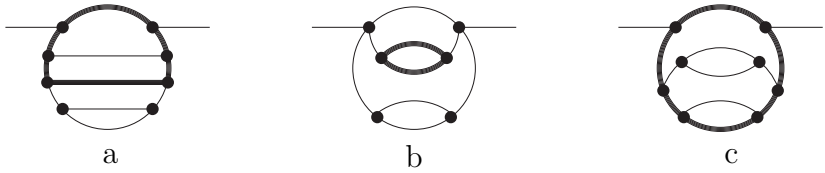
\begin{figure}[hbt]
\begin{center} \begin{picture}(360,50)(0,0)
\SetScale{0.77}
\Line(20,50)(48,50) \CArc(70,30)(30,0,360)
\CArc(70,30)(30.5,-10,190) \CArc(70,30)(29.5,-10,190)
\CArc(70,30)(31,-10,190) \CArc(70,30)(29,-10,190)
\Line(92,50)(120,50) \Line(41,36)(99,36) \Line(40,23)(100,23)
\Line(40,23.5)(100,23.5) \Line(40,22.5)(100,22.5)
\Line(40,24)(100,24) \Line(40,22)(100,22)
\Line(48,10)(92,10) \Vertex(48,50){3} \Vertex(92,50){3}
\Vertex(41,36){3} \Vertex(99,36){3} \Vertex(40.5,23){3}
\Vertex(99.5,23){3} \Vertex(48,10){3} \Vertex(92,10){3}
\Text(55,-10)[]{a}

\Line(170,50)(198,50) \CArc(220,30)(30,0,360)
\CArc(220,20)(22,40,140)
\CArc(220,20)(22.5,40,140)\CArc(220,20)(21.5,40,140)
\CArc(220,20)(23,40,140)\CArc(220,20)(21,40,140)
\CArc(220,50)(22,180,360)
\CArc(220,50)(22.5,220,320) \CArc(220,50)(21.5,220,320)
\CArc(220,50)(23,220,320) \CArc(220,50)(21,220,320)
\CArc(220,-16)(30,50,130) \Line(242,50)(270,50) \Vertex(198,50){3}
\Vertex(242,50){3} \Vertex(204,35){3} \Vertex(236,35){3}
\Vertex(238,6){3} \Vertex(202,6){3} \Text(170,-10)[]{b}

\Line(320,50)(348,50) \CArc(370,30)(30,0,360)
\CArc(370,30)(30.5,0,360) \CArc(370,30)(29.5,0,360)
\CArc(370,30)(31,0,360) \CArc(370,30)(29,0,360)
\CArc(370,10)(30,20,160) \CArc(370,56)(30,230,310)
\CArc(370,-16)(30,50,130) \Line(392,50)(420,50) \Vertex(348,50){3}
\Vertex(392,50){3} \Vertex(342,19){3} \Vertex(398,19){3}
\Vertex(351,33){3} \Vertex(388,6){3} \Vertex(389,33){3}
\Vertex(352,6){3} \Text(290,-10)[]{c}
\end{picture}
\end{center}
\caption{The component diagrams corresponding to
  Fig.\ref{15} and giving contribution to ${[\beta_{m^2}]^i_j}^{(4)}$}
\end{figure}
The single poles of these diagrams are

\begin{center}
\begin{tabular}{|c|c|c|}\hline
 & &  \\[-3mm]
a & b & c  \\[1mm]
 \hline
  & &  \\
$\;\;\displaystyle{\frac{3}{8\epsilon}}\;\;$&
$\;\;\displaystyle{-\frac{1}{24\epsilon}}\;\;$&
$\;\;\displaystyle{\frac{1}{48\epsilon}}\;\;$\\
& &  \\
\hline
\end{tabular}
\end{center}

After subtraction of these diagrams from the corresponding
superfield ones and replacing the Yukawa vertices with the external
lines by a hypercharge, we get after reduction of tensor
structures
\begin{eqnarray}
\Delta\beta_\xi^{(4)}=\frac{17}{6}F^{(A\bar A)}_1
+\frac{7}{3}F^{(A\bar A)}_2+\frac{1}{2}\tilde{F}^{(A\bar A)}_2
+\frac{7}{6}F^{(A\bar A)}_3+\frac{1}{8}F^{(A\bar A)}_4
+\frac{7}{3}F^{(A\bar A)}_5+\frac{1}{2}\tilde{F}^{A\bar
A}_5\nonumber\\+\left(2\zeta (3)-\frac{7}{6}\right)F^{(A\bar A)}_6
-\left(\zeta (3)-\frac{3}{4}\right)F^{(A\bar A)}_7-\left(\zeta
(3)-\frac{2}{3}\right)F^{(A\bar A)}_8 +\frac{1}{4}F^{(A\bar A)}_9 ,
\end{eqnarray}
where
$$F^{(A\bar A)}_1=
{A} ^{jkl}{{\cal Y}_{j}^{i}\bar A} _{ikm}y^{mnp}\bar
y_{lnq}y^{qst}\bar y_{pst}\ \ \ \ \
F^{(A\bar A)}_2=
{A} ^{jkl}{{\cal Y}_{j}^{i}\bar y} _{ikm}y^{mnp}\bar
A_{lnq}y^{qst}\bar y_{pst}$$
$${\tilde F}^{(A\bar A)}_2=
{A} ^{ikl}{\bar y} _{ikm}y^{mnp}\bar A_{lnq}y^{qst}{\cal
Y}_{s}^{j}\bar y_{jpt}\ \ \ \ \
F^{(A\bar A)}_3=
{A} ^{ikl}{\bar y} _{ikm}y^{jmp}{\cal Y}_{j}^{n}\bar
y_{lnq}y^{qst}\bar A_{pst}
$$
$$F^{(A\bar A)}_4=
{y}^{ikl}{\bar y} _{ikm}A^{mjp}{\cal Y}_{j}^{n}\bar
A_{lnq}y^{qst}\bar y_{pst}\ \ \ \ \
F^{(A\bar A)}_5=
{y} ^{jkl}{{\cal Y}_{j}^{i}\bar A} _{ikm}A^{mnp}\bar
y_{lnq}y^{qst}\bar y_{pst}$$
$${\tilde F}^{(A\bar A)}_5=
{y} ^{ikl}{\bar A} _{ikm}A^{mnp}\bar y_{lnq}y^{qst}{\cal
Y}_{s}^{j}\bar y_{jpt}\ \ \ \ \
F^{(A\bar A)}_6=
{A} ^{jkl}{{\cal Y}_{j}^{i}\bar A} _{ikm}y^{mnp}\bar
y_{npq}y^{qst}\bar y_{lst}$$
$$F^{(A\bar A)}_7=
{y} ^{ikl}{\bar A} _{ikm}{\cal Y}_{j}^{m}A^{jnp}\bar
y_{npq}y^{qst}\bar y_{lst}\ \ \ \ \
F^{(A\bar A)}_8=
{y} ^{ikl}{\bar A} _{ikm}{\cal Y}_{j}^{m}y^{jnp}\bar
y_{npq}A^{qst}\bar y_{lst}$$
$$F^{(A\bar A)}_9=
{A} ^{ikl}{\bar A} _{ikm}y^{jmp}{\cal Y}_{j}^{n}\bar
y_{lnq}y^{qst}\bar y_{pst}
$$
Notice that a direct calculation of the tadpoles in components
requires the evaluation of nearly thirty different diagrams. The
same proportion is valid for the other four loop contributions to
the D-term renormalization.

\section{Conclusion}
We have found that all the information about the renormalizations
of the soft SUSY breaking terms in the N=1 SUSY gauge theory is
contained in a rigid, unbroken theory. In the case of the non-Abelian
gauge group, the RG equations for the soft terms are obtained from
the anomalous dimensions of the matter and vector superfields by
acting of the differential operators~\cite{AKK,KV,JJ}. 
In the presence of the Abelian
gauge group, to calculate the renormalization of an additional 
Fayet-Iliopoulos term, one needs an analysis of superfield
diagrams. To find the contribution proportional to the soft scalar
mass $(m^2)^i_j$ (the square of gaugino mass $M \bar M$), one
needs to take the self-energy diagrams for the vector superfield
and replace one of the external vertices with the hypercharge ${\cal
Y}^i_j$ by $(m^2)^i_j$ ($M \bar M \delta^i_j$). In this case, there
is no need to do any calculations except in superfields.

The other contributions (proportional to $A\bar A$ and $M \bar A$)
can be found from the analysis of the matter superfield propagator
diagrams in a rigid theory and the corresponding component
diagrams in a softly broken theory extracting from the latter  the
contribution of the tadpole graphs.
In this case, one needs to calculate additionally some component
diagrams the number of which is essentially reduced compared to a
direct component calculation.

\section*{Appendix}

Throughout the paper we use the standard metric $g_{\mu
\nu}=diag(1,-1,-1,-1)$. In this metric the chiral matter
superfield and the vector superfield can be written as
\begin{eqnarray}
\Phi&=&\phi+i\,\theta\sigma^\mu
\bar\theta\partial_\mu\phi-\frac14
\theta\theta\bar\theta\bar\theta
\partial^\mu\partial_\mu\phi+\sqrt{2}\theta\psi+\frac{i}{\sqrt{2}}
\theta\theta\bar\theta\bar\sigma^\mu\partial_\mu\psi+
\theta\theta F,\\
V&=&C+i\,\theta\chi-i\,\bar\theta\bar\chi+\frac{i}{2}\theta\theta
N-\frac{i}{2}\bar\theta\bar\theta \bar N
-\theta\sigma^\mu\bar\theta
v_\mu+i\theta\theta\bar\theta\bar\lambda-\frac12\theta\theta\bar\theta
\bar\sigma^\mu\partial_\mu\chi-i\bar\theta\bar\theta\theta\lambda
\nonumber\\&+&
\frac12\bar\theta\bar\theta\theta\sigma^\mu\partial_\mu\bar\chi
+\frac12\theta\theta\bar\theta\bar\theta
D-\frac14\theta\theta\bar\theta\bar\theta \partial^\mu\partial_\mu
C.
\end{eqnarray}
The interaction of a matter superfield with an Abelian superfield
is given by eq.(\ref{rigidlag}) and for the triple vertex has the
form (in arbitrary gauge, not the Wess-Zumino one)
\begin{eqnarray}
\int d^2\theta d^2\bar{\theta} ~\bar{\Phi}^i (e^{V})^j_i\Phi_j& =&
\frac{i}{\sqrt{2}}
\bar\phi\lambda\psi-\frac{i}{\sqrt{2}}\bar\psi\bar\lambda\phi+
\frac12\bar\phi D\phi -\frac{i}{2}\bar\phi\, v_\mu
\partial^\mu\phi
+\frac{i}{2}\partial^\mu\bar\phi\, v_\mu \phi+\frac12\bar\psi\bar\sigma^\mu v_\mu\psi\nonumber\\
&+&\partial^\mu\bar\phi\, C \partial^\mu\phi
-\frac12\bar\phi\,\partial^\mu\partial_\mu
C\phi-\frac{1}{\sqrt{2}}\bar\phi\bar\chi
\bar\sigma^\mu\partial_\mu\psi-\frac{1}{\sqrt{2}}\partial_\mu\bar\psi\bar\sigma^\mu
\chi\phi+\frac{i}{\sqrt{2}}\bar\psi\bar\chi F\nonumber\\&-&
\frac{i}{\sqrt{2}}\bar F\chi\psi+\bar F C F+\frac{i}{2}\bar F N
\phi-\frac{i}{2}\bar \phi \bar N F -\frac{i}{2}\bar \psi\, C
\bar \sigma^\mu\partial_\mu\psi+\frac{i}{2}
\partial_\mu\bar\psi C\bar\sigma^\mu\psi\nonumber.
\end{eqnarray}
In the Wess-Zumino gauge only the first line is left. However, for
our analysis one needs all the vertices and therefore some new
diagrams arise.\vspace{0.8cm}

{\Large \bf Acknowledgements}\vspace{0.7cm}

The authors would like to thank M.Tentyukov for his help in using
DIANA program and A.Kotikov, S.Mikhailov and A.Bakulev for
valuable discussions. Financial support from RFBR grants \#
99-02-16650 and \# 00-15-96691 is kindly acknowledged.

\end{document}